\documentclass[]{jfm}
\usepackage{graphicx}
\usepackage{epstopdf,epsfig}
\usepackage{newtxtext}
\usepackage{newtxmath}
\usepackage{natbib}
\usepackage{hyperref}
\hypersetup{
    colorlinks = true,
    urlcolor   = blue,
    citecolor  = black,
}

\newcommand{\RomanNumeralCaps}[1]
\linenumbers
\usepackage{siunitx}
\usepackage{ctable}
\usepackage{threeparttable}
\usepackage{diagbox}
\usepackage{pifont}

\title{Boundary compliance selects heterogeneous dynamics in shear-thickening suspensions}

\author{Li-Xin Shi\aff{1},
  Meng-Fei Hu\aff{2}\and Song-Chuan Zhao\aff{2,3}\corresp{\email{songchuan.zhao@outlook.com}}}

\affiliation{
\aff{1}Key Laboratory of Advanced Engineering Materials and Structural Mechanical Behavior and Intelligent Control for Universities in Hunan Province, School of Civil and Environmental Engineering, Changsha University of Science and Technology, Changsha 410114, China\\
\aff{2}State Key Laboratory for Strength and Vibration of Mechanical Structures, School of Aerospace Engineering, Xi'an Jiaotong University, Xi'an 710049, China\\
\aff{3}Department of Physics, Kyushu University, Fukuooka, Japan}

\begin{document}
\maketitle

\begin{abstract}
The mechanical properties of confining boundaries can fundamentally alter the flow behaviour of shear-thickening suspensions. We study a dense cornstarch suspension sheared beneath a viscous silicone-oil layer, using the oil viscosity to tune boundary compliance. Flow visualisation and rheometry reveal two distinct regimes. With compliant boundaries, long-lived heterogeneities emerge via density waves or persistent clusters, maintained by a balance between interface deformation and particle rearrangement. With more resistant confinement, we observe transient jamming events, marked by abrupt spanning of load-bearing structures across the suspension thickness and the emergence of secondary stress waves. The onset stress of these events remains constant at the DST threshold, independent of bounding viscosity. Our results reveal that boundary compliance selects the lifetime and morphology of heterogeneous structures, offering a means to amplify otherwise short-lived microscopic processes and providing new insight into the interplay between shear thickening, shear jamming, and confinement mechanics.
\end{abstract}


{\bf MSC Codes }  {\it(Optional)} Please enter your MSC Codes here


\section{Introduction}
\label{sec:headings}

Dense suspensions composed of micron-sized, non-Brownian solid particles immersed in a viscous fluid are known to exhibit shear thickening. This characteristic is marked by a rise in viscosity with an increase in the applied stress, $\tau$, or strain rate, $\dot{\gamma}$~\citep{guazzelli2018rheology,morris2020shear}. Recent studies have associated this behavior with the formation of frictional contacts when the applied stress exceeds a characteristic threshold, $\tau^{*}$, which is determined by interparticle repulsion~\citep{wyart2014discontinuous,clavaud2017revealing,seto2013discontinuous}. The constitutive curve of such a suspension typically displays an S-shape, indicating a transition from a low-viscosity, lubricated state at low stresses to a high-viscosity, frictional state at high stresses. Despite this average observation in experiments, the flow is not steady. The thickening process is characterized by significant fluctuations in bulk stress or strain rate~\citep{lootens2003giant,hermes2016unsteady}. Recent studies with spatially-resolved experiments have discovered considerable spatiotemporal structures linked to these fluctuations~\citep{ovarlez2020density,saint2018uncovering,rathee2017localized}. These findings challenge the current understanding of the shear-thickening phenomenon and call for a more nuanced interpretation.

The inherent instability of the uniform flow in a thickened state, as indicated by the S-shaped constitutive curve, leads to the formation, percolation, and collapse of localized contact networks of particles. Numerical simulations have confirmed this underlying mechanism~\citep{nabizadeh2022structure,van2023minimally,Goyal2024}. Experimental observations have revealed intermittent, spatially localized jamming events~\citep{rathee2020localized,rathee2022structure}, which may be interpreted as macroscopic manifestations of contact-network percolation. These processes are fundamental to the response of a shear-thickened state to external drivings. The importance of frictional contacts between the particles suggests a shared framework that includes both shear thickening and shear jamming~\citep{peters2016direct,Brown_2014,han2019stress}. Boundary confinement is thus deemed critical to the evolution of the contact network in the flow~\citep{brown2012role}. However, previous studies have largely overlooked the potential impact of boundary compliance on the thickening behavior~\citep{Boundaryconditions2024}, which is a substantial limitation not only for understanding shear-thickening but also for controlling its dynamics in numerous natural and industrial settings where soft boundaries or free surfaces are commonly encountered.

In this study, we explore the influence of boundary compliance on the evolution of heterogeneities in shear-thickening suspensions. Our findings reveal distinct regimes of dynamics; compliant boundaries lead to persistent heterogeneity due to dynamic instability, while resistant confinement results in the emergence of transient, load-bearing clusters through localized spanning events.

\section {Experimental protocol}

\begin{figure*}
	\centering
	\includegraphics[scale=1]{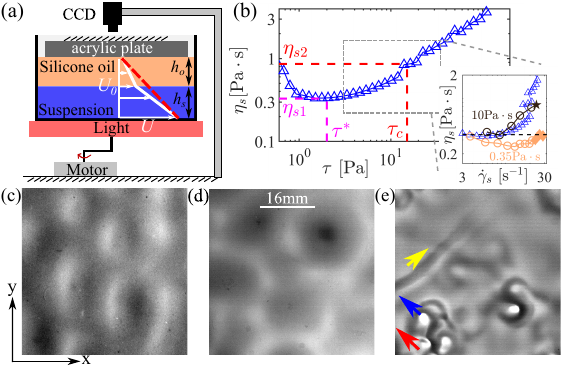}
	\caption{(a) A schematic of the experimental setup. (b) Rheogram of the dense cornstarch suspension for $\Phi=0.42$. For detailed description, see the text. 
		(c-e) Snapshots of three different states of inhomogeneity (see Movies S1, S2, and S3 for multimedia views). The red arrows indicate the instantaneous drive direction. The blue arrow refers to the instantaneous directions of high-$\phi$ regions' motion. The yellow one indicates the propagating event in the vorticity direction. Inset of panel (b): Comparison of independently measured rheological curves, $\eta_s(\dot{\gamma}_s)$, and results obtained in our experiments with silicone oils of different viscosities.}
	\label{fig:device}
	
\end{figure*}

In order to investigate the influence of boundary compliance on flow instabilities, we design an experiment utilizing a stratified shear cell, which was driven by orbital oscillations of the bottom at a moving speed denoted as $U$. This specific setup was chosen to avoid the stress gradients that commonly occur in the vorticity direction in parallel-plate rheometry. As depicted in figure~\ref{fig:device}(a), a thin layer of silicone oil was introduced between the suspension and the fixed top plate, which has a diameter of $110~\si{mm}$. The thickness of this oil layer, denoted as $h_o$, is set to $2~\si{mm}$, while the thickness of the suspension, denoted as $h_s$, is set to $3~\si{mm}$. The local thickness of suspension can be measured on demand via a laser displacement sensor. By altering the oil viscosity $\eta_o$, we realize a smooth transition from weak to strong interfacial constraints. 

A key quantity in our experiment is the interfacial velocity field $\vec{U_0}(x,y)$, measured using particle image velocimetry (PIV) with $50~\si{\mu m}$ hollow tracer particles dispersed at the oil-suspension interface. These particles were tracked using particle image velocimetry (PIV). The amplitude of the mean interfacial velocity, $\bar{U_0}$, was obtained by averaging over a measurement area of $25~\si{cm^2}$ and one oscillation period. 
The averaged shear rates in the suspension and oil layers were then calculated as $\dot{\gamma}_s=(U - \bar{U_0})/h_s$ and $\dot{\gamma}_o = \bar{U_0}/h_o$ respectively. Under low-Reynolds-number conditions, the averaged shear stress at the suspension-oil interface is determined by $\tau=\eta_o \dot{\gamma}_o$, and the apparent suspension viscosity, $\eta_s=\tau /\dot{\gamma}_s$, was derived from the continuity of the shear stress $\tau$ across the interface. However, when local quantities are of interest (e.g. the local stress field or $\dot{\gamma}_s$ under heterogeneous states), the corresponding local velocity field $\vec{U_0}(x,y)$ is used instead.

The suspension is prepared by suspending cornstarch particles (Aladdin S116030) in deionized water, with caesium chloride added to achieve density matching. These dry starch particles have a mass density of $\rho_p=1.5\si{kg/m^3}$ and an average diameter of $d_p=15~\si{\mu m}$. The volume fraction is maintained at $\Phi=0.42$. The suspension was illuminated from beneath by a monochrome LED panel, with inhomogeneous spatial transmission of light indicating uneven density, as demonstrated by \citet{Shi_Hu_Zhao_2024}. Specifically, darker regions correspond to higher density of particles (as shown in Figs.~\ref{fig:device}(c-d)). Therefore, the onset and evolution of heterogeneity can be quantitatively characterized by the spatial standard deviation of the transmission light, as shown in  Appendix~\ref{appA}. Rheological measurements are carried out separately using a parallel-plate rheometer (Anton Paar MCR 302) in stress-controlled mode (figure\ref{fig:device}(b)). These measurements indicated that shear thickening starts at $\tau^{*}=2~\si{Pa}$ corresponding  to an apparent suspension viscosity of $\eta_{s1}=0.33~\si{Pa \cdot s}$. The onset of Discontinuous Shear Thickening (DST), accompanied with pronounced fluctuations in the bulk strain rate, occurs at $\tau_c=15~\si{Pa}$, corresponding to an apparent suspension viscosity of $\eta_{s2}=0.86~\si{Pa \cdot s}$. These two characteristic viscosities define the critical boundary compliance, which will be discussed in further detail later.

\section {Results}
We first take an overview of the experimental observations. As $\dot{\gamma}_s$ increases, the suspension exihibits heterogeneities in density (see Appendix~\ref{appA}). Depending on the viscosity of the bounding silicone oil $\eta_o$, we identify three distinct heterogeneous flow states (\textit{state 1–3}), as illustrated in figure.~\ref{fig:device}(c–e), within two regimes.

\begin{figure*}
	\centering
	\includegraphics[scale=1]{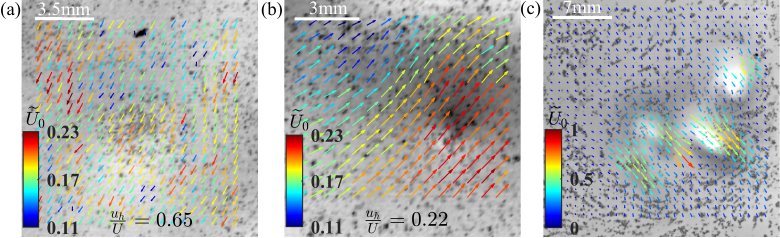}
	\caption{(a-c) Local flow fields corresponding to \textit{state~1}, \textit{state~2} and \textit{state~3}, respectively. Note that $\widetilde{U}_0=\frac{\lvert\vec{U}_0(x,y)\rvert}{U}$ refers the scaled interface velocity, and $u_h$ denotes the speed of the heterogeneous structure.}
	\label{fig:flowfield}
	
\end{figure*}

For relatively low viscosity, where $\eta_o <\eta_{s2}$, the suspension develops inhomogeneities as shear thickening progresses. The emergence of these nonuniformities is accompanied by the deformation of oil-suspension interface, with a typical peak-to-valley height difference of $\Delta h \approx 0.5~\si{mm}$. These inhomogeneities travel along the mainstream and self-organize into a hexagonal pattern [figure~\ref{fig:device}(c,d)]. Such patterns can persist throughout the entire experimental duration (Movies S1 and S2). In contrast, as $\eta_o>\eta_{s2}$, the heterogeneous dynamics are more violent, often causing substantial density gradients upstream (figure~\ref{fig:flowfield}c) and interface deformation ($\Delta h \approx 0.8~\si{mm}$). In this regime, high-density regions move almost in phase with the container and are frequently followed by secondary propagating events along the vorticity direction as seen in  figure~\ref{fig:device}e and Movie S3. The differing persistence of the inhomogeneities in these states can be quantified by the variance of the backlight transmission ratio and the time series of interface deformation (see Appendix~\ref{appA}).

Further insight comes from the local interface flow fields, as shown in figure~\ref{fig:flowfield} (a-c), which reveal a clear distinction among these heterogeneous states. In \textit{state~1}, the tracer particles velocity remain lower than that of the heterogeneous structures, suggesting that the motion of the inhomogeneity propagate as kinematic density waves~\citep{Shi_Hu_Zhao_2024}.
In contrast, for \textit{state~2},  the tracer particles move synchronously with the high-density region (figure~\ref{fig:flowfield}(b)), as if forming locally jammed clusters~\citep{Shi_Zhao2024}. 
While the exact dimensions of the clusters remain under investigation, they likely do not span the entire shear-gradient (vertical) direction. One indication is their slower motion compared to the bottom driving velocity $U$. This feature contrasts sharply with \textit{state~3}, where, once formed, the structures move in phase with the driving. As shown in figure~\ref{fig:flowfield}c, in \textit{state~3}, the amplitude of the scaled interfacial velocity $\widetilde{U}_0$ (where $\widetilde{U}_0=\lvert\vec{U_0}(x,y)\rvert/U$) approach 1. This indicates that they are locally jammed and extend the gap thickness. As they evolve, low-density tails develop upstream and $\widetilde{U}_0$ decreases, marking the progressive dissolve of these structures.

The origin of these heterogeneous states requires further investigation. 
In figure~\ref{fig:phase_diagram}(a) we plot the average stress, $\tau$, versus the shear rate, $\dot{\gamma}_s$. Here, $\dot{\gamma}_s$ is nondimensionalized using the particle Reynolds number, $\mathrm{Re}_p=\rho_p \dot{\gamma} d_p^{2}/\eta_f$.  The observed $\mathrm{Re}_p\ll 1$ indicates negligable inertia effect. Nevertheless, due to the density-dependent constitutive relation, instabilities can be raised by the kinematic accumulation. This type of role of particle inertia in the emergence of unsteady flow in particle suspensions has been documented~\citep{Johri2002,duru2002constitutive}. 
As $\mathrm{Re}_p$ ($\dot{\gamma}_s$) increases, the suspension progress towards the shear-thickening regime. As demonstrated in figure~\ref{fig:phase_diagram}(a), the critical particle Reynolds number for the onset of instability increases with the viscosity of the bounding silicone oil, $\eta_o$. A similar trend is observed for the corresponding onset stress, $\tau_{on}$. This trend persists within the regime of persistent heterogeneous states, but saturates once $\eta_o$ exceeds $\eta_{s2}$, beyond which unsteady flow is triggered at a constant stress $\tau_c$, corresponding to the onset of discontinuous shear thickening (DST). The non-monotonic dependence of the instability onset on $\eta_o$, illustrated in figure~\ref{fig:phase_diagram}(d), suggests the existence of distinct underlying mechanisms.  In the following, we propose a minimal model to understand these boundary effects.

\section {Discussions}
\begin{figure}
	\centerline{\includegraphics[scale=1]{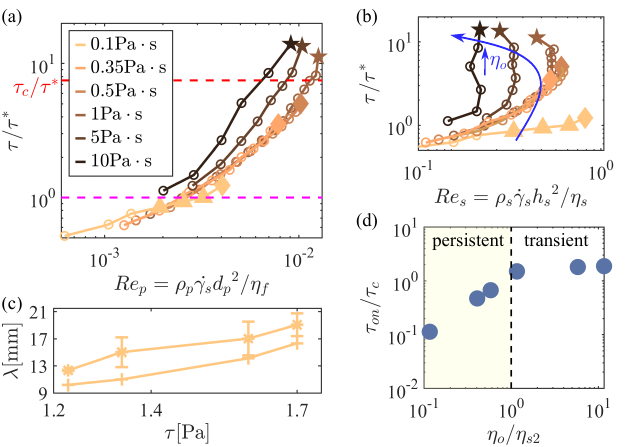}}
	\caption{(a) State diagram illustrating distinct unsteady regimes observed at varying silicone oil viscosities and their relation to the particle Reynolds number. Symbols represent different dynamical states: $\circ$ represents the uniform state, $\blacktriangle$ denotes \textit{state 1}, $\blacklozenge$ indicates \textit{state 2}, and  $\bigstar$ corresponds to \textit{state 3}. (b) Normalized shear stress plotted against the suspension Reynolds number, with data and symbols consistent with those in (a). (c) Comparison of wavelengths $\lambda$ (defined as the distance between adjacent low-density regions along the mainstream): $+$ indicates the theoretical prediction from Eq.~\ref*{eq.verticle_balance}, while $\ast$ shows experimentally measured values. Note that the shear stress $\tau$ is varied by adjusting the thickness of the silicone oil layer while keeping the driving velocity fixed. (d) Variation of onset stress with silicone oil viscosity.}
	\label{fig:phase_diagram}
\end{figure}

These non-uniform flow modes are related to shear-thickening and are fundamentally different from that in Newtonian viscosity-stratified systems, where instability can occur at vanishing Reynolds numbers~\citep{Yih_1967,Hooper_Boyd_1987}. In contrast, as shown in figure~\ref{fig:phase_diagram}(b), the onset Reynolds number for inhomogeneity in suspensions exhibits a non-monotonic dependence on $\eta_o$. Experiments using Newtonian viscosity-stratified fluids do not produce similar instabilities, further highlighting the unique nature of the observed phenomena. In the shear-thickening suspension, the instability is governed by the competition between the kinematic accumulation and pressure-driven migration of the particle phase \citep{Shi_Hu_Zhao_2024,Shi_Zhao2024}. Kinematic accumulation promotes the growth of particle clusters, leading to a buildup of stress. The buildup of stress are not necessary aligned with the size growth of the force network~\citep{van2023minimally}. As the contact network percolates, a sudden rise in local stress may lead to the collapse of the contact network~\citep{Goyal2024,van2023minimally}. The magnitude of stress that the boundary can withstand without deformation determines the actual dynamics~\citep{Boundaryconditions2024}. The apparent viscosity of the inhomogeneous states, $\eta_s$, are maintained at various levels of thickening (cf. figure~\ref{fig:device}b inset). To gain a deeper understanding of the non-uniform states, we first examine the regime of $\eta_o<\eta_{s2}$, where long-lived heterogeneities are observed. 

\subsection {Long-lived inhomogeneity for $\eta_{o}<\eta_{s2}$}
For $\eta_o<\eta_{s1}$, as shown in figure~\ref{fig:phase_diagram}a, a stress around $\tau^{*}$ triggers the emergence of flow inhomogeneity, which aligns with the onset of shear thickening. In this regime, the system initially exhibits \textit{state~1}. As $\dot{\gamma}_s$ increases, \textit{state~1} transforms into \textit{state~2}. Conversely, for $\eta_o>\eta_{s1}$, the onset of inhomogeneous flow shifts to higher stresses beyond $\tau^*$ and directly manifests as \textit{state~2}, bypassing the intermediate \textit{state~1}.

This variation in flow modes with bounding viscosity can be explained by two characteristic timescales: $t_1=1/{\dot{\gamma}_o}$, the time for interface deformation, and $t_2=1/ \dot{\gamma}_s$, the time taken for particles to pass each other. The ratio of these timescales defines a dimensionless parameter, 
\begin{equation}
	\epsilon=\frac{t_1}{t_2}=\frac{\dot{\gamma}_s}{\dot{\gamma}_o}=\frac{\eta_o}{\eta_s},
	\label{eq.epsi}
\end{equation}
where the last equality arises from continuity of shear stress across the interface. The dilation of the particle phase can occur if the interface deforms faster than particles move past each other, i.e., $\epsilon \le 1$.

In the case of $\eta_o < \eta_{s1}$, the condition $\epsilon < 1$ is always satisfied. Consequently, the stress buildup from growing inhomogeneities is promptly relaxed, preventing it from exceeding $\tau^*$ (the threshold for shear thickening and the onset of frictional contacts). Particle contacts remain transient, and the inhomogeneity manifests as \textit{state~1}. The transiency of contacts makes the inhomogeneity the only dynamic mode that shows no hystersis~\citep{Shi_Zhao2024}. Since the viscous resistance of the oil layer (approximately $\tau$) alone cannot balance the particle pressure $\sigma_N \sim \mathcal{O}(\tau^*)$, capillary pressure arising from interface deformation must be considered. This pressure scales as $2\Delta h \Gamma / (\lambda/2)^2$, where $\lambda$ is the wavelength defined as the distance between adjacent low-density regions along the mainstream, $\Gamma = 20~\si{mN/m}$ is the interfacial tension, and $\Delta h \approx 0.5~\si{mm}$ is the measured deformation amplitude. The resulting stress balance gives:
\begin{equation}
	\sigma_N \sim \mathcal{O}(\tau^*) = \frac{2\Delta h}{(\lambda/2)^2} \Gamma + \tau.
	\label{eq.verticle_balance}
\end{equation}
This relationship neglects the vertical fluid drag during particle dilation (i.e., Darcy-type flow)~\citep{jerome2016unifying,athani2022transient,athani2025transients}. One reason for this omission is that the viscosity of the silicone oil exceeds that of the solvent (water) by more than two orders of magnitude. Another is that, in the present experimental configuration, fluid migration is expected to occur primarily within the flow-vorticity plane. Equation~\eqref{eq.verticle_balance} implies that the wavelength $\lambda$ adjusts with increasing $\tau$ to maintain the transient contact state. Experimental measurements support this prediction, showing a monotonic increase of $\lambda$ with shear stress $\tau$ (figure~\ref{fig:phase_diagram}(c)).

\begin{figure}
	\centerline{\includegraphics{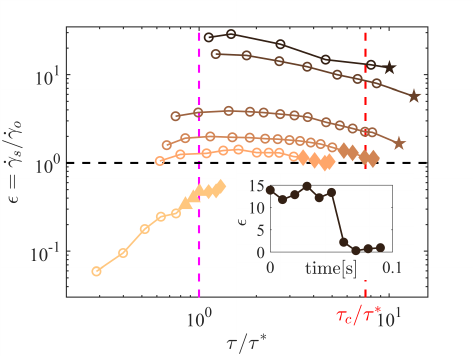}}
	\caption{The characteristics of heterogeneous in different value of the dimensionless number $\epsilon=\dot{\gamma}_s/\dot{\gamma}_o$. The data and symbols here are consistent with those in figure~\ref{fig:phase_diagram}. Inset: A discontinuous drop in local $\epsilon$ for $\eta_{o}=10~\si{Pa\cdot s}$. Note that, to determine $\epsilon$ locally, the averaged velocity was measured within a fixed $1~\si{mm}^2$ area.}
	\label{fig:time_scale}
\end{figure}

On the other hand, when $\eta_o > \eta_{s1}$, the onset of shear thickening ($\tau\approx\tau^*$ and $\eta_s\approx\eta_{s1}$) yields $\epsilon>1$. In this case, dilation of the particle phase is obstructed, aligning the dynamics with those of a rigid boundary. The growth of the contact network is hindered by the vertical confinement and shear profile and is likely limited to submacroscopic scale~\citep{Goyal2024,van2023minimally}. These immature contact structures are not readily observable via transmission contrast in our experiments. As $\tau$ increases, shear thickening develops, and $\eta_s$ rises, while the oil viscosity $\eta_o$ remains constant. This causes the dimensionless parameter $\epsilon$ to decrease. As $\epsilon$ approaches 1, the dilation constraint eases, potentially allowing the structures to grow to larger sizes without being abruptly destroyed by a sudden rise in stress. A balance between influx and efflux could eventually be reached, thereby  maintaining the inhomogeneous pattern. Since a more viscous bounding requires deeper shear-thickening to achieve $\epsilon=1$, the onset stress for the inhomogeneity, $\tau_{on}$, is observed to shift to higher values with increasing $\eta_o$ (see figure~\ref{fig:phase_diagram}d). Furthermore, as the dilation constraints are relaxed, further shear thickening becomes unsustainable under low-Reynolds-number conditions. Consequently, the effective suspension viscosity asymptotically approaches that of the bounding oil (see figure~\ref{fig:device}(b) inset), and the time scale ratio $\epsilon$ stabilizes near 1, exhibiting minimal variation with increasing stress $\tau$ (see figure~\ref{fig:time_scale}).

We clarify two aspects of the long-lived inhomogeneous states discussed above. First, such persistent structures in \textit{state~2} can also emerge from the \textit{state~1} regime ($\eta_o < \eta_{s1}$) upon further increasing $\dot{\gamma}$ and $\tau$ (see figure~\ref{fig:phase_diagram}a and \ref{fig:time_scale}). However, this evolution involves highly nonlinear dynamics~\citep{Shi_Zhao2024}, distinct from that emerge directly from the uniform state when $\eta_o > \eta_{s1}$. We therefore do not elaborate on this path here. Second, the instability mechanism underlying these inhomogeneous states is fundamentally different from that in Newtonian stratified flows, where the interfacial instability is driven by discontinuities in viscosity and shear profile~\citep{Yih_1967,Hooper_Boyd_1987}. In contrast, the inhomogeneity here arises when the viscosities of the two layers become comparable (\textit{i.e.}, $\epsilon \to 1$), corresponding to a transition in the shear profile from the solid white to the dashed red line in figure~\ref{fig:device}(a).

\subsection {Transient jamming for $\eta_{o}>\eta_{s2}$}

As previously discussed, the onset stress of inhomogeneity, $\tau_{on}$, exhibits a positive correlation with $\eta_{o}$. This correlation persists until $\eta_{o}$ surpasses $\eta_{s2}$, at which point the dynamics enter a distinct regime. This new regime is characterized by several unique features. Notably, the viscosity threshold $\eta_{s2}$ corresponds to a stress scale of $\tau_c=15~\si{Pa\cdot s}$. As per our prior discussion, it is expected that $\tau_{on}>\tau_c$ in this regime, and that $\tau_{on}$ increases with $\eta_{o}$ to ensure $\epsilon=\eta_{o}/\eta_s$ decreases to 1. However, when $\eta_{o}$ exceeds $\eta_{s2}$, the onset of heterogeneity remains constant at $\tau_{on}\approx \tau_c$, regardless of $\eta_o$ (refer to figures~\ref{fig:phase_diagram}d). Additionally, despite the globally averaged $\epsilon$ remaining larger than 1, locally the heterogeneity is associated with an abrupt, transient drop to $\epsilon \sim 0$ (figure~\ref{fig:time_scale}, inset). This indicates that the local shear rate is temporarily reduced to 0, consistent with the formation of a localized load-bearing cluster spanning the gap, or a transient jamming event. 

\begin{figure}
	\centerline{\includegraphics{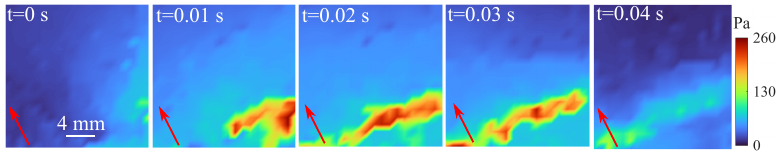}}
	\caption{Snapshot of the stress field of a transient event along the vorticity direction. The red arrow indicate the instantaneous drive direction. The stress field was derived from the measured local interfacial velocity field, $\vec{U}(x,y)$, obtained via PIV analysis. In the present experiment, $\eta_{o} = \SI{10}{Pa~s^{-1}}$ and the average shear rate is 38 \si{s^{-1}}.}
	\label{fig:stressmap}
\end{figure}

Simulations~\citep{nabizadeh2022structure,van2023minimally,Goyal2024} and experiments~\citep{Heterogeneous2025,kim2023stress} have both demonstrated that the intensification of shear-thickening is underpinned by the growth of the contact network in both size and strength. At high $\tau$, these clusters percolate across the depth of the suspension but promptly collapse due to their intense interaction with the confining boundaries. Similarly, in our experiments, as the load-bearing clusters emerge and expand to approach the full gap thickness, they significantly drag the bounding silicone oil, resulting in a sharp increase in local stress. The resulting stress gradient, typically on the order of $100~\si{Pa~mm^{-1}}$ (as shown in figure~\ref{fig:stressmap}), drives cluster rearrangement and eventual breakup. Their lifetimes do not conform to a single characteristic timescale but are typically comparable to one oscillation cycle (see Movie S3). As local transient jamming occurs, \textit{i.e.}, $\dot{\gamma}_s\to 0$ locally, the amplitude of local stress is primarily proportional to the viscosity of the bounding oil $\eta_o$. Thus, a higher $\eta_o$ amplifies the stress gradient, accelerating the cluster collapse. Under rigid confinement, their lifetimes become significantly shorter than a single oscillation cycle~\citep{Boundaryconditions2024}.

The collapse of spanned load-bearing clusters can further trigger secondary stress wave propagation. The collapse may be incomplete, breaking into mesoscopic scale rather than particle scale. Consequently, mesoscopic perturbations induced during breakup may rapidly grow by linking pre-existing force chains dispersed throughout the suspension. This behavior echoes dynamic jamming fronts observed in wide-gap Couette cell and in impact experiments~\citep{peters2016direct,han2019stress}, although along the vorticity direction rather than the driving or gradient directions. The propagation velocity reaches approximately $1.3~\si{m~s^{-1}}$ (see figure~\ref{fig:stressmap}), nearly an order of magnitude higher than the imposed drive speed. It is important to realize that the compliant bounding is critical to both the emergence and the lifetime of this secondary stress wave. Under the viscous boundary conditions studied here, $\eta_{o}>\eta_{s2}$, a characteristic time for their survival is comparable to that defined by the average shear rate (e.g., $\approx 0.03~\si{s}$ in figure~\ref{fig:stressmap}), whereas they are rarely visible under rigid confinement~\citep{rathee2022structure,Moghimi2025}.

\subsection {Further remarks}
The understanding of shear thickening, particularly discontinuous shear thickening (DST), has undergone a paradigm shift in recent years. The emerging view emphasizes the load-bearing network of frictional contacts. The macroscopic inhomogeneous states described in this study may be interpreted as various evolutionary stages of force network in shear thickening-from the onset of frictional contacts (\textit{state 1}, $\tau\sim \tau^{*}$), through the formation of localized clusters (\textit{state 2}, $\tau > \tau^{*}$), to their spanning the boundary gap (\textit{state 3}, $\tau > \tau_c$). These states correspond to successive stages of viscosity proliferation (see apparent viscosity $\eta_s$ in figure~\ref{fig:device}b), which, in the light of the nature of DST, may reflect different stages of stress transmission and contact-network development. The emergence and spanning of localized load-bearing structures do not necessarily require density inhomogeneites on macroscopic scales. Rather, their presence could promote kinematic accumulation and amplify density inhomogeneity~\citep{Shi_Hu_Zhao_2024}. The macroscopic inhomogeneity observed here arises from the prolonged lifetime of these structures under compliant boundaries. During the thickening process, stress promotes the growth of load-bearing structures but also leads to their yielding. Boundary compliance provides a feedback mechanism that determines the stage at which a stress-activated structure becomes sufficiently strong to deform the confinement. 
We speculate that this feedback prolongs the survival of force chains and allows microscale structural dynamics to manifest as macroscopic flow nonuniformity. Otherwise, the microscopic structures are quite short-lived, they remain too weak to be detected by techniques such as boundary stress microscopy (BSM). As a result, the suspension appears relatively uniform in experiments for $\tau<\tau_c$. It is only when $\tau > \tau_c$ that significant macroscopic stress inhomogeneity is triggered~\citep{rathee2017localized,rathee2022structure}. 

Recent simulations further refine the evolution of force networks in shear-thickening phenomena. A close link between DST and shear jamming (SJ) has been established~\citep{peters2016direct,Moghimi2025}. Nevertheless, these two transitions remain distinct: DST is characterized by intermittent resistance during flow, whereas SJ is marked by robust, system-spanning arrest. Consequently, differences in the evolution of the force/contact network are expected. Beyond $\tau^*$, force chains can connect into networks of system size. As $\tau$ increases, locally over-constrained or mechanically rigid clusters begin to form within these contact networks. 
Specifically, \cite{Goyal2024} identified over-constrained clusters based on their coordination numbers. There, the percolation of an over-constraint network was proposed as the origin of DST criticality and as the mechanism responsible for significant stress fluctuations. Our experiments are consistent with this picture in several respects. The transient events of $\dot{\gamma}_s(\epsilon)\rightarrow 0$ observed for $\eta_{o} > \eta_{s2}$ suggest that the abrupt emergence of effectively spanning load-bearing clusters is marked by $\tau_c$, the onset stress of DST. The correlation between these events and stress fluctuations provides further support for this interpretation. This transition is robust against variations in $\eta_{o}$, reinforcing its stress-activated nature. The secondary stress waves observed following the failure of these clusters further suggest the presence of abundant mesoscopic load-bearing clusters during DST, consistent with the criticality argument. At the same time, the transient nature of the spanning events highlights the fragility of the underlying force networks. A more stringent definition of mechanical rigidity based on the pebble game shows that the contact network at DST is dominated by sliding contacts, which, despite a relatively high coordination number, do not provide sufficient constraints for mechanical rigidity~\citep{van2023minimally, Santra2025}. Mechanically rigid structures percolate only at stresses far beyond DST, constituting another regime in the flow-state diagram.  How to interpret these results in systems with large aspect ratios and macroscopic inhomogeneities, as well as the relationship between these two percolation transitions, remains an open question. Under rigid confinement, spanning events are too small and short-lived to be clearly resolved. The viscous confinement protocol used here provides an alternative means to enhance the observability of these processes.

The role of boundary compliance emphasized here has broader practical implications. In industrial and natural processes, suspensions are often enclosed by composite confinements with weak points. Even in pipe flow, the outlet acts as an open boundary causing shock waves in stress~\citep{bougouin2024frictional}. As demonstrated here and by \citet{Boundaryconditions2024}, deformability tends to concentrate high-density and high-stress regions while reducing the overall degree of thickening in the bulk. Strategies such as incorporating viscoelastic patterns might be employed to regulate shear-thickening behavior. It is worthy noting that even the weakest elastic confinement tested by \citet{Boundaryconditions2024} permits no long-lived heterogeneity. Therefore, future research should carefully distinguish between deformability arising from elasticity and that from viscosity.

\section{Conclusion}

The experiments presented here demonstrate that the compliance of the confining boundary plays a critical role in controlling the form and persistence of heterogeneous flow in shear-thickening suspensions. By systematically varying the viscosity of the bounding layer, we identified two distinct regimes corresponding to distinct thickening stages. Compliant confinement prolongs the lifetime of load-bearing clusters, allowing them to manifest as persistent, macroscopic density patterns, whereas more resistant confinement promotes transient jamming events with rapid collapse. The dimensionless timescale ratio $\epsilon = \eta_o / \eta_s$ governs the observed transition between these regimes, while the onset of transient jamming events, where load-bearing structures effectively span the system, occurs at a constant stress scale, independent of the bounding viscosity. These observations potentially link boundary deformation mechanics to the stress-driven growth and yielding of frictional contact structures, aligning with the view that the percolation of over-constrained clusters coincides with the onset of discontinuous shear thickening. Beyond the fundamental insight, this study suggests strategies for manipulating suspension behavior in natural and industrial flows. Moreover, the approach also offers a means to reveal short-lived microscopic processes that remain inaccessible under rigid confinement, opening new directions for research on the interplay between suspension microstructure, flow instabilities, and confinement mechanics.

\backsection[Supplementary data]{\label{SupMat}Supplementary movies are available at \\https://doi.org/10.1017/jfm.2019...Movies illustrating the movement of density waves, clusters, and transient fluctuations can be found in \textbf{S1\_wave.mp4}, \textbf{S2\_cluster.mp4}, and \textbf{S3\_fluctuation.mp4}, respectively.}

\backsection[Funding]{This project is supported by the National Natural Science Foundation of China (Grant No. 12172277). L. X. Shi is additionally supported by Innovative projects of Key Disciplines of Civil Engineering of Changsha University and Science and Technology (25ZDXK05).}


\backsection[Declaration of interests]{The authors report no conflict of interest.}


\backsection[Author ORCIDs]{S. C. Zhao, https://orcid.org/0000-0001-6790-9663; L. X. Shi, https://orcid.org/0009-0003-8605-7648; M. F. Hu, https://orcid.org/0009-0007-5469-6516}

\appendix
\section{Detection of flow heterogeneity via transmitted light intensity}\label{appA}
\begin{figure}
	\centerline{\includegraphics{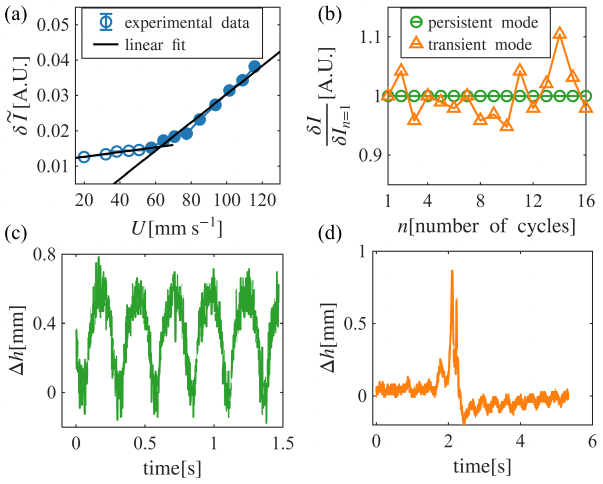}}
	\caption{(a) Spatial standard deviation of the transmitted light intensity as a function of the driving velocity $U$. Open circles indicate the homogeneous state, whereas filled circles correspond to the heterogeneous state. The critical point is determined from the intersection of two fitted branches of $\delta \widetilde{I}(U)$. (b) Time sequences of $\delta I_t$ for persistent (green) and transient (orange) inhomogeneities. The values are averaged over each oscillation cycle. (c) Time series of interface deformation for the persistent heterogeneities. (d) Time series of interface deformation for the transient heterogeneities. }
	\label{fig:light_intensity}
\end{figure}

Our previous experiments have shown that the transmitted-light ratio $I_t(x,y)$ provides a measure of heterogeneity and its temporal evolution in the suspension~\citep{Shi_Zhao2024}, where the subscript $t$ denotes time. Though a quantitative relation between $I_t(x,y)$ and the particle density distribution can be established, we use this observable semi-quantitatively here. At a given time, the spatial standard deviation, $\delta I_t$, reflects the magnitude of density heterogeinities. The average of $\delta I_t$ over three oscillation cycles, denoted as $\delta \widetilde{I}$, is used to characterize the heterogeneous states. Figure~\ref{fig:light_intensity}(a) shows the variation of $\delta\widetilde{I}$ as a function of the driving velocity $U$. The critical point, which separates the uniform and heterogeneous states, is defined by the intersection of two linearly fitted branches of $\delta \widetilde{I}(U)$. In the uniform regime, the gradual increase of $\delta\widetilde{I}$ is associated with the sidewall effects. Beyond the critical point, whereas the rapid increase in $\delta\widetilde{I}$ is dominanted by the spontaneous bulk heterogeneity. 

The heterogeneous states exhibit two distinct types: persistent and transient. Figure~\ref{fig:light_intensity}b shows representative $\delta I_t$ evolutions of these two types.  In the persistent mode (green curve), the standard deviation remains constant over successive oscillation cycles $n$. Whereas in the transient mode (orange curve), it exhibits pronounced fluctuations, reflecting the intermittent formation and dissolution of heterogeneous patterns. The autocorrelation of $\delta I_t$ can thus be used to define the lifetime of transient heterogeneities.
This distinction is further corroborated by the time series of interface deformation at a fixed position. As shown in figure~\ref{fig:light_intensity}(c, d), a persistent heterogeneous state leads to periodic interfacial deformation (panel c), whereas a transient event results in only a short-lived deformation (panel d). 



\bibliographystyle{jfm}
\bibliography{wave_JFM}

\end{document}